\documentclass{ws-jai}
\usepackage[flushleft]{threeparttable}
\usepackage{hyperref}
\usepackage{xcolor}
\hypersetup{
    colorlinks,
    linkcolor={red!50!black},
    citecolor={blue!50!black},
    urlcolor={blue!80!black}
}

\begin{document}

\catchline{}{}{}{}{} 

\markboth{Brajesh Kumar}{First light preparations of the 4m ILMT}

\title{First light preparations of the 4m ILMT}

\author{Brajesh Kumar$^{1}$, Hitesh Kumar$^{1}$, Khushal Singh Dangwal$^{1}$, Himanshu Rawat$^{1}$, Kuntal Misra$^{1}$, Vibhore Negi$^{1,2}$, Mukesh Kumar Jaiswar$^{1}$, Naveen Dukiya$^{1}$, Bhavya Ailawadhi$^{1,2}$, Paul Hickson$^{3}$ and Jean Surdej$^{4, 5}$}

\address{
$^{1}$Aryabhatta Research Institute of Observational Sciences, Manora Peak, Nainital - 263001, India.\\
$^{2}$Deen Dayal Upadhyaya Gorakhpur University, Gorakhpur 273009, India.\\
$^{3}$Department of Physics and Astronomy, University of British Columbia, 6224 Agricultural Road, Vancouver, BC V6T 1Z1, Canada.\\
$^{4}$Institute of Astrophysics and Geophysics, University of Li\`ege, All\'ee du 6 Ao\^ut 19c, 4000 Li\`ege, Belgium.\\
$^{5}$Astronomical Observatory Institute, Faculty of Physics, Adam Mickiewicz University, ul. Sloneczna 36, 60-286 Poznan, Poland.\\
}

\maketitle

\corres{$^{1}$brajesh@aries.res.in, brajesharies@gmail.com}

\begin{history}
\received{(to be inserted by publisher)};
\revised{(to be inserted by publisher)};
\accepted{(to be inserted by publisher)};
\end{history}

\begin{abstract}
The 4m International Liquid Mirror Telescope (ILMT) is a zenith-pointing optical observing facility at ARIES Devasthal observatory (Uttarakhand, India). The first light preparatory activities of the ILMT were accomplished in April 2022 followed by on-sky tests that were carried out at the beginning of May 2022. This telescope will perform a multi-band optical (SDSS $g’$, $r’$ and $i’$) imaging of a narrow strip ($\sim$22$'$) of sky utilizing the time-delayed integration technique. 
Single-scan ILMT images have an integration time of 102 sec and consecutive-night images can be co-added to further improve the signal-to-noise ratio. An image subtraction technique will also be applied to the nightly recorded observations in order to detect transients, objects exhibiting variations in flux or position. Presently, the facility is in the commissioning phase and regular operation will commence in October 2022, after the monsoon. This paper presents a discussion of the main preparation activities before first light, along with preliminary results obtained.
\end{abstract}


\section{Introduction}
Operation of liquid mirror telescopes (LMTs) is very different from that of  conventional glass mirror telescopes. As evident from the name, the primary mirror of a LMT is formed by a thin layer of reflecting liquid (usually mercury). The surface of the mercury-filled container takes the shape of a paraboloid when rotated around a vertical axis at a constant rate. This results from a balance between the gravitational and centrifugal forces. LMTs do not track mechanically. However, image motion due to the rotation of the Earth can be compensated using a CCD operating in time-delayed integration (TDI) mode.  Normally, TDI images are degraded by star-trail curvature and differential drift rates across the CCD. However, this can be mitigated by an optical corrector that is designed to compensate for these effects  \citep[see][]{Gibson-1992,Hickson-1998}.
The single-scan integration time is equal to the time taken for a target to cross the detector. Notably, as the same region of sky will pass over the telescope with an offset 4 min in right ascension, the nightly images in different bands can be co-added to reach fainter magnitudes. It has been already demonstrated that diffraction-limited images can be achieved with LMTs. The cost-effectiveness, optimal seeing position observations, easy maintenance, etc. make them an efficient astronomical tool \citep{Borra-1982, Borra-1989, Gibson-1991, Hickson-1994ApJ, Hickson-1994}. The largest LMT that has been constructed so far is the Large Zenith Telescope \citep{Hickson-2007LZT}.

\begin{figure}
\centering
\includegraphics[scale=0.2]{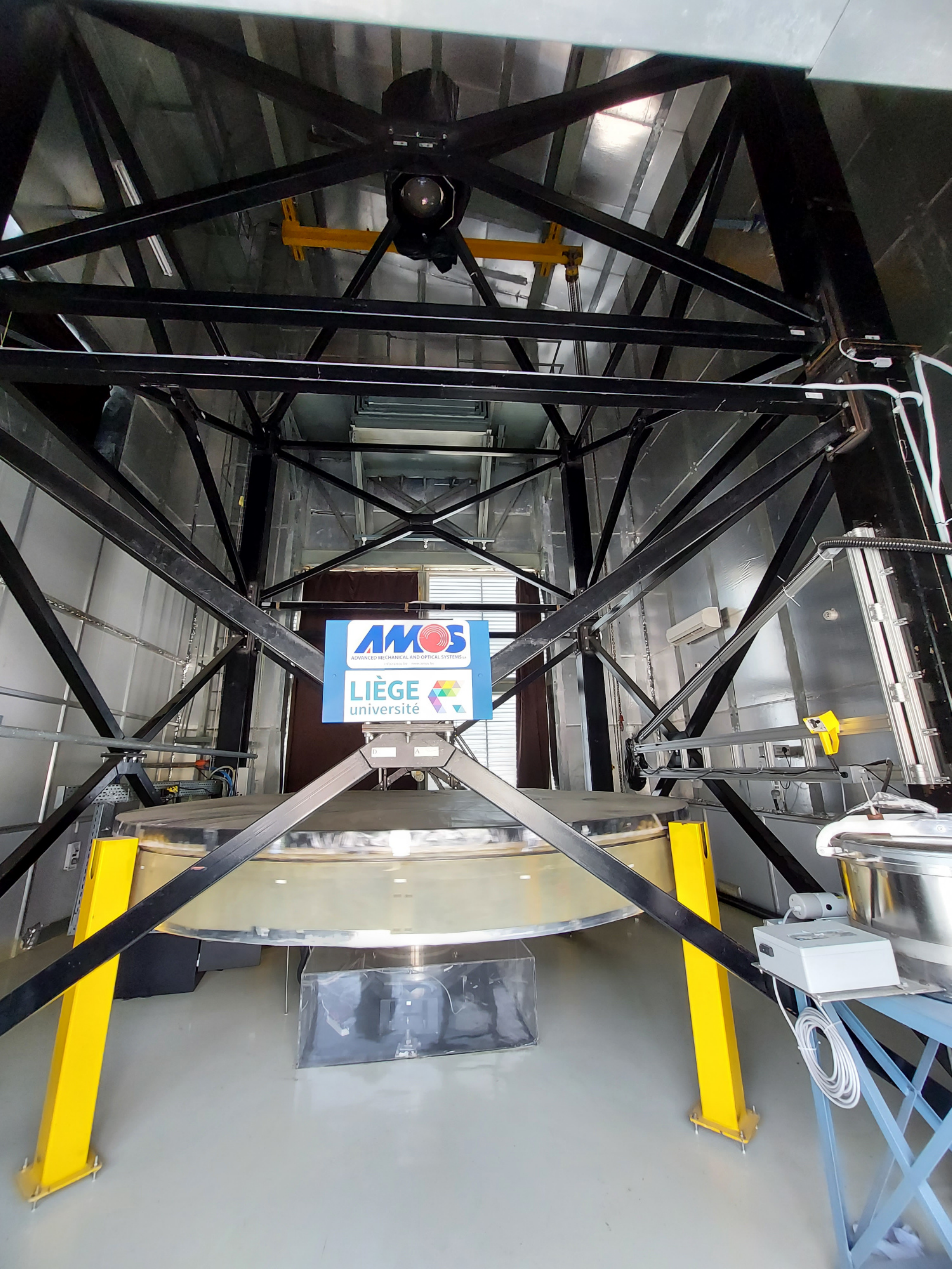}
\caption{Front view of the ILMT facility. The mirror (covered with a mylar film) is sitting over the air-bearing (inside the plexiglass box). A mercury storage stainless steel reservoir is fixed on top of a metallic frame (blue). The mercury pumping system is attached to the right side black pillar. Two safety pillars (yellow) are also visible in the front. The optical corrector and CCD camera assembly can be seen at the top.}
\label{ILMT-front}
\end{figure}

The 4m International Liquid Mirror Telescope (ILMT) is the first LMT designed specifically for astronomical observations. It is located at Devasthal observatory (79$^{\circ}$ 41$'$ 07$''$ E, +29$^{\circ}$ 21$'$ 41$''$) at an altitude of 2450 m in the central Himalayan region  near Nainital, India, a site that offers a dark sky and good seeing \cite{Sagar-2000}. Two conventional astronomical telescopes, the 1.3-m Devasthal Fast Optical Telescope (DFOT) and the 3.6-m Devasthal Optical Telescope (DOT) are already operational there \citep{Sagar-2012, Sagar-2019, Brij-2018}. The ILMT is a 4-m zenith-pointing telescope that consists of several essential components including an air bearing, primary mirror, a prime-focus corrector, CCD assembly and telescope structure \citep[see][and references therein]{Surdej-2018, Kumar-2014Thesis, Kumar-2018MN, 2022JApA...43...10K}. The effective focal length of the optical system is 9.44 m. 

As shown in Fig. ~\ref{ILMT-front}, The f/2 primary mirror is made of carbon fiber-epoxy skin over a closed-cell foam core. Twelve radial ribs support a dish-shaped top. The dish supports a thin polyurethane surface that is formed by spin-casting to a  parabolic shape. It is this surface that supports a thin layer (approximately 3 mm) of mercury. The mirror is mounted on an air-bearing (Kugler model RT-600T), which suppresses vibrations. This bearing can support a load of approximately 1000 kg and is itself supported by a three-point mount that allows precise leveling of the bearing. 

The air bearing is supplied with dry filtered air, provided by two air compressors, a pneumatic air control system, membrane air dryers, filters, and pressure, temperature, humidity and dew-point sensors. This system removes particulates larger than 10 nm, and lowers the dew point of the air to typically $-60^\circ$ C. 

The ILMT prime-focus assembly includes a $4096 \times 4096$ pixel shutter-less CCD camera (different parameters are provided in Table~\ref{tab_ILMT}), a filter slide, and a five-lens TDI-compensating optical corrector. Motorized mechanisms allow remote control of focus, tilt, camera rotation, and filter position.

\begin{table}
\small
\centering
\caption{Specifications of the ILMT detector.}\label{tab_ILMT}
\begin{tabular}{ll}
\hline
Parameter                    &  Value                     \\ \hline
Array size                   &  4096 $\times$ 4096 pixels \\
Pixel size                   &  $0.33''$ pixel$^{-1}$     \\
Readout noise                &  5.0 e$^{-1}$              \\
Gain                         &  4.0 e$^{-1}$/ADU          \\
Integration time (TDI)       &  102 sec                   \\
\hline
\end{tabular}
\end{table}

Each of the components and instruments was verified and calibrated at various stages of manufacturing and installation. However, it was decided to perform various preparatory activities before the first light as described in Section~\ref{prep}.

\section{First-light preparatory activities}\label{prep}
In this section, we present the major preparatory activities which took place before the first sky tests of the ILMT. It includes leveling of the bowl (container), azimuth alignment of the primary mirror with respect to the air bearing, vertical runout verification of the bowl, mylar installation, and mirror formation. These are described below:

\subsection{Leveling of the air bearing}\label{mirror-align}
\begin{figure}
\centering
\includegraphics[scale=0.7]{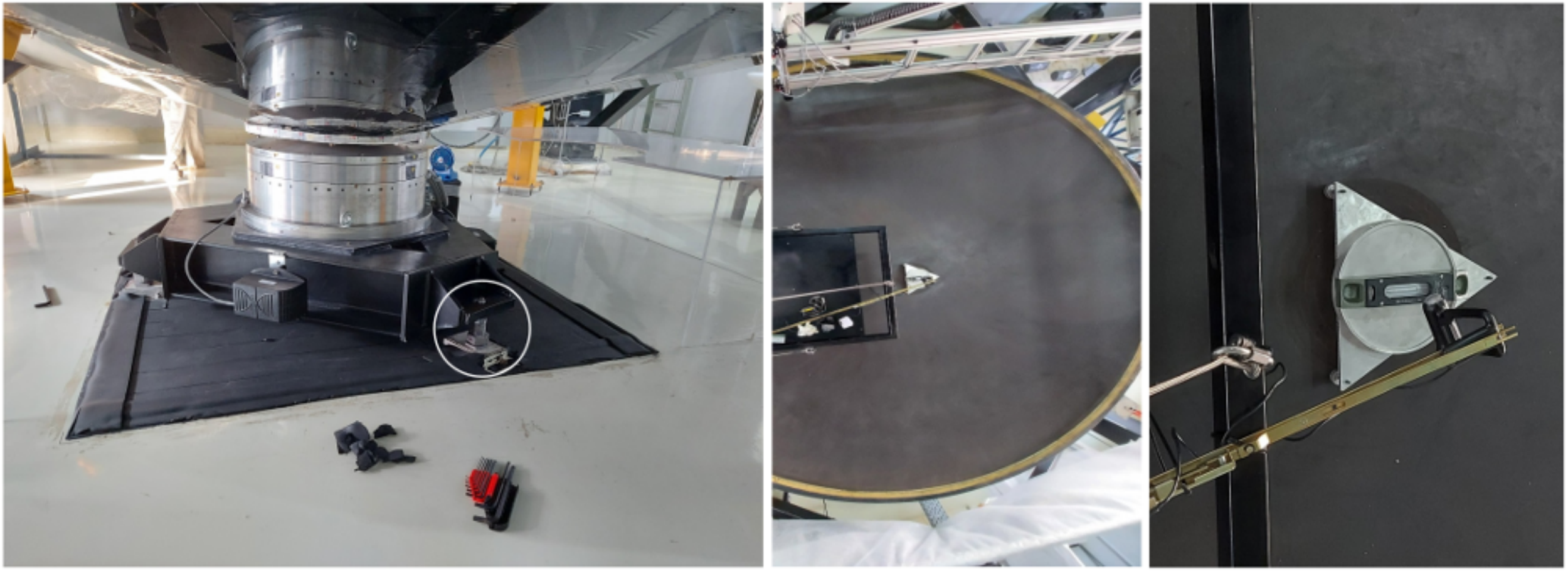}
\caption{Experimental setup for the mirror level measurement. Left panel: The ILMT air-bearing sitting over the three-point mounting. The encircled region is indicating one of the three screws that was used to adjust the bubble level (see, Section~\ref{mirror-align}). Middle panel: The spirit level installed over the metallic plates is positioned at the center of the bowl. Right panel: A zoomed section showing the bubble level. The images were saved with a camera each time after rotating the bowl.}
\label{ILMT-MA-1}
\end{figure}

\begin{figure}
\centering
\includegraphics[width=\columnwidth]{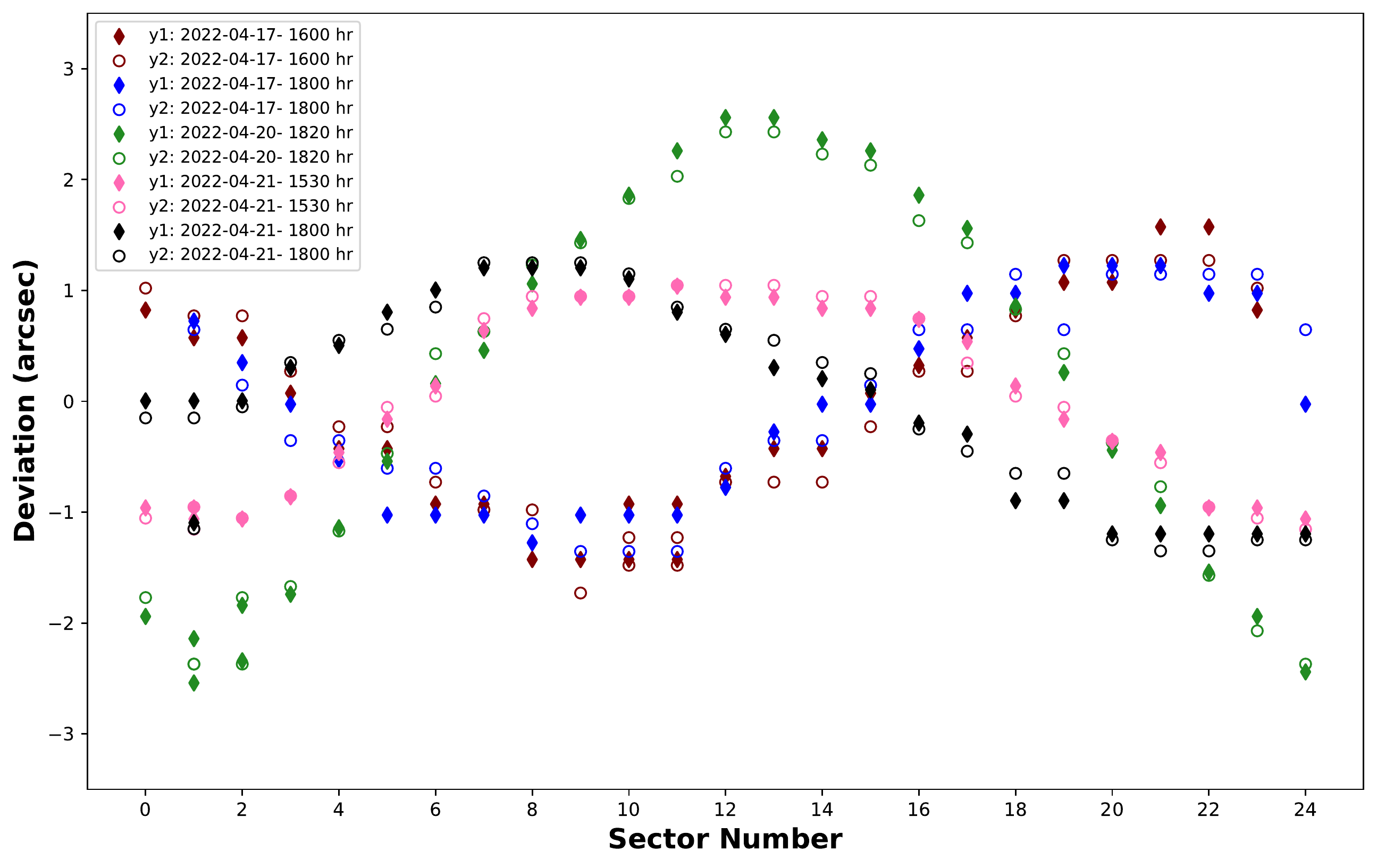}
\caption{Deviations at different sectors are plotted. Here, Y1 and Y2 indicate the positions of the bubble level on opposite sides. The date of measurement and time are indicated in the legend.}
\label{ILMT-MA-plot}
\end{figure}

The rotation axis of the mirror must be vertical to within a fraction of an arcsec. Otherwise, a diamond-ring like structure might result in the stellar images \citep{Hickson-2006}. Therefore, mirror leveling must be critically verified. This activity should be repeated whenever the mirror is stopped for the cleaning and maintenance purposes. The ILMT air bearing has a three-point mounting system for leveling (marked with A, B and C). The mirror has been divided into 24 sectors ($\sim$15$^{\circ}$ each), each labeled with a number (1, 2, ..., 24).
The tools used to level the bowl are: a high-accuracy spirit level (least count = 2 arcsec), one triangular support frame and one flat circular metallic plate. The air-bearing and triangular support frame can be independently adjusted with the help of three screws. First, the spirit level position was set in the center by adjusting the triangular plate screws (Fig.~\ref{ILMT-MA-1}, middle and right panels).
Then, the air-bearing three-point mounting screws were further used to adjust the spirit level position (Fig.~\ref{ILMT-MA-1}, left panel). The bubble positions were noted for each sector position of the bowl. The alignment procedure was repeated during different times of the day and the results are plotted in Fig~\ref{ILMT-MA-plot}. It is evident that the mirror rotation axis may be off by 2 to 3 arcsec. Variation has been observed during different times of the day. Deformation of the telescope pier due to temperature gradients could be the possible cause behind such variations. Due to time constraints, we proceeded for the first light. However, we expected that increased accuracy can be achieved, and this will be explored in the future.  

\subsection{Air bearing and primary mirror azimuth alignment}
Dry compressed air is supplied to the ILMT air bearing for its operation. This air is supplied by two air compressors that are installed in an auxiliary  building. The air-bearing is insulated with an acrylic enclosure and foam adhesive tapes to maintain an ambient temperature of $\sim$20$^{\circ}$ C. 

In order to protect the air bearing from damage in case of a mercury imbalance, the primary mirror is not firmly attached to the air bearing. Rather, there is an interface between two precisely-ground stainless-steel plates, one attached to the mirror and the other to the air bearing. The plates are constrained radially by a central post, but are otherwise free to move if an imbalance occurs. In order to minimize vertical run-out of the polyurethane surface, the two plates must be aligned correctly in azimuth. To facilitate this, there are X+/X+, X--/X--, Y+/Y+ and Y--/Y-- marks on the air bearing plate and on the plate attached to the mirror. Alignment was performed by closing the valve that allows air to flow to the bearing and then locking the bearing internal rotation using three screws. The bowl was then lifted up, with the help of four carjacks. The above marks were then manually aligned after unlocking the screws and resupplying the air. Again, the air was cut off, and bearing plates were locked. The bowl was then carefully lowered into position, and the air supply was turned on.        

\begin{figure}
\centering
\includegraphics[width=\columnwidth]{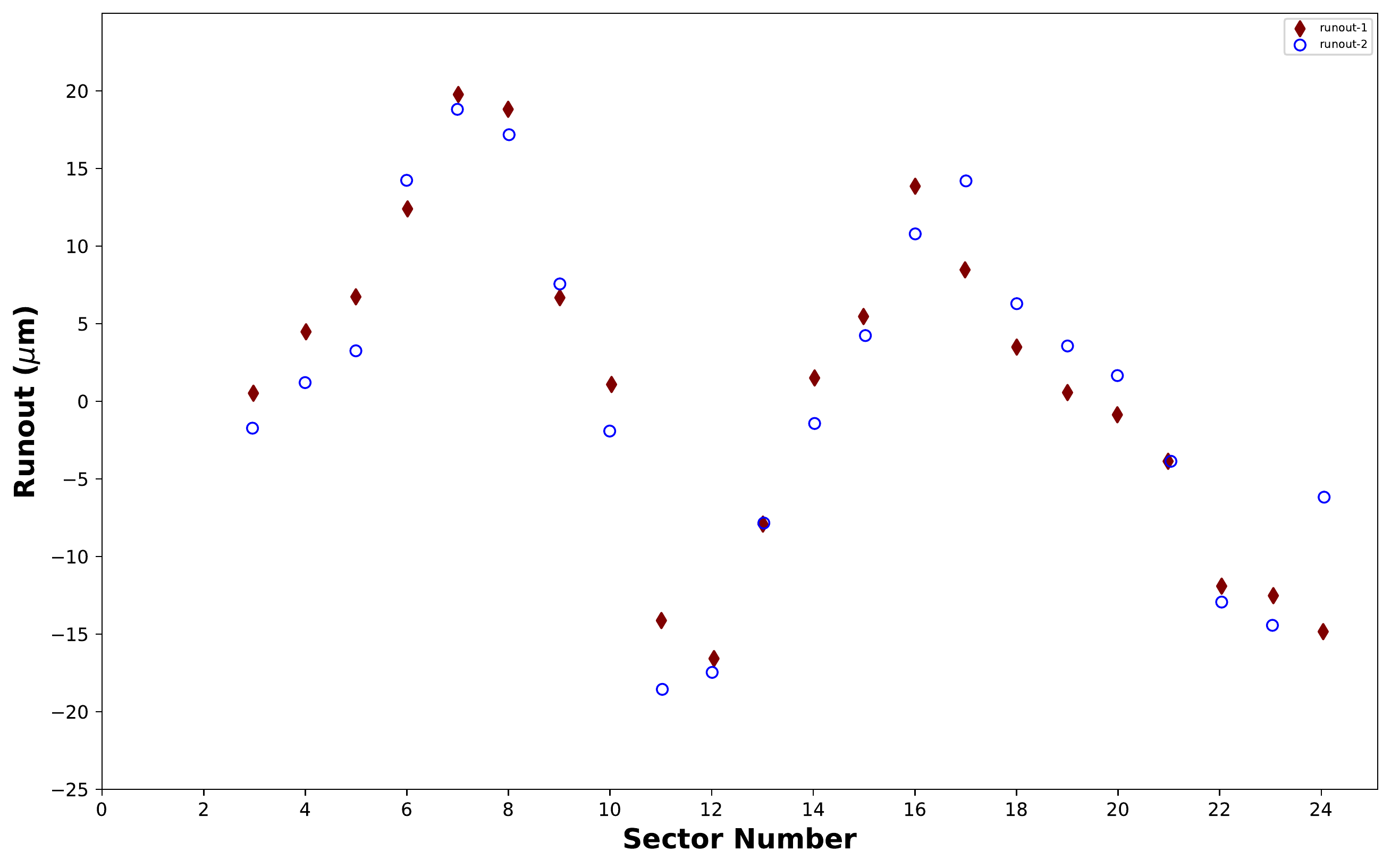}
\caption{The vertical run-out is plotted against sector number. Two independent measurements are indicated with different symbols.}
\label{ILMT-vertical}
\end{figure}

\subsection{Verification of mirror vertical runout}
Vertical runout measurements of the mirror were performed using two dial gauges. These were installed on two of the four safety pillars that are installed close to the mirror rim, and multiple measurements at each sector number (1--24) were performed. The runout variation is plotted in Figure~\ref{ILMT-vertical}. It can be seen that the runout deviation is $\pm$20 microns for both sets of measurements, which is within the acceptable range.

\subsection{Mylar installation and mirror formation}

Tests with previous LMTs indicate that for mirror diameters greater than about 2 m, turbulence develops in the boundary layer between the rotating mirror and the surrounding air. Vortices that form in the boundary layer cause pressure fluctuations at the surface of the mercury. These fluctuations raise waves on the surface, with amplitudes that can exceed 1 $\mu$m. Such waves cause unacceptable image degradation and must be mitigated. The only reliable method to overcome this problem found to date is to cover the mirror with a film of optical-quality mylar \citep{Hickson-2007LZT}. 

The mylar film is stretched over the mirror, above the mercury, and co-rotates with it. This protects the mercury from the vortices, which now occur above the mylar film. Consequently, the image quality is greatly improved. Furthermore, the mylar film also suppresses harmful mercury vapour that is produced during the first few hours of operation. The ILMT uses a scientific grade mylar film which has a thickness of $\sim 1.2~\mu$m. As the width of available rolls of mylar is less than the diameter of the mirror, several pieces of mylar were attached together using adhesive tape in order to cover the entire mirror. To aid in installation, a stand was constructed that holds the completed cover wound up in a horizontal roll. The stand was then positioned beside the mirror so that the cover could be unwound, pulling it over the mirror. 

Prior to installation of the mylar cover, the polyethylene surface of the mirror was thoroughly cleaned using isopropanol. A mylar support net was then prepared using several thin nylon strings. Initially, five nylon strings were installed over the bowl, parallel to sectors 24--12, 22--10, 20--8, 18--6 and 16--4. Then, three more nylon strings were installed perpendicular to them. Mylar film was carefully unrolled from the mylar stand and spread across the mirror. Adhesive tape was used to secure the mylar all around the periphery. Three additional nylon strings were installed over the mylar to further secure it.

\begin{figure}
\centering
\includegraphics[width=\columnwidth]{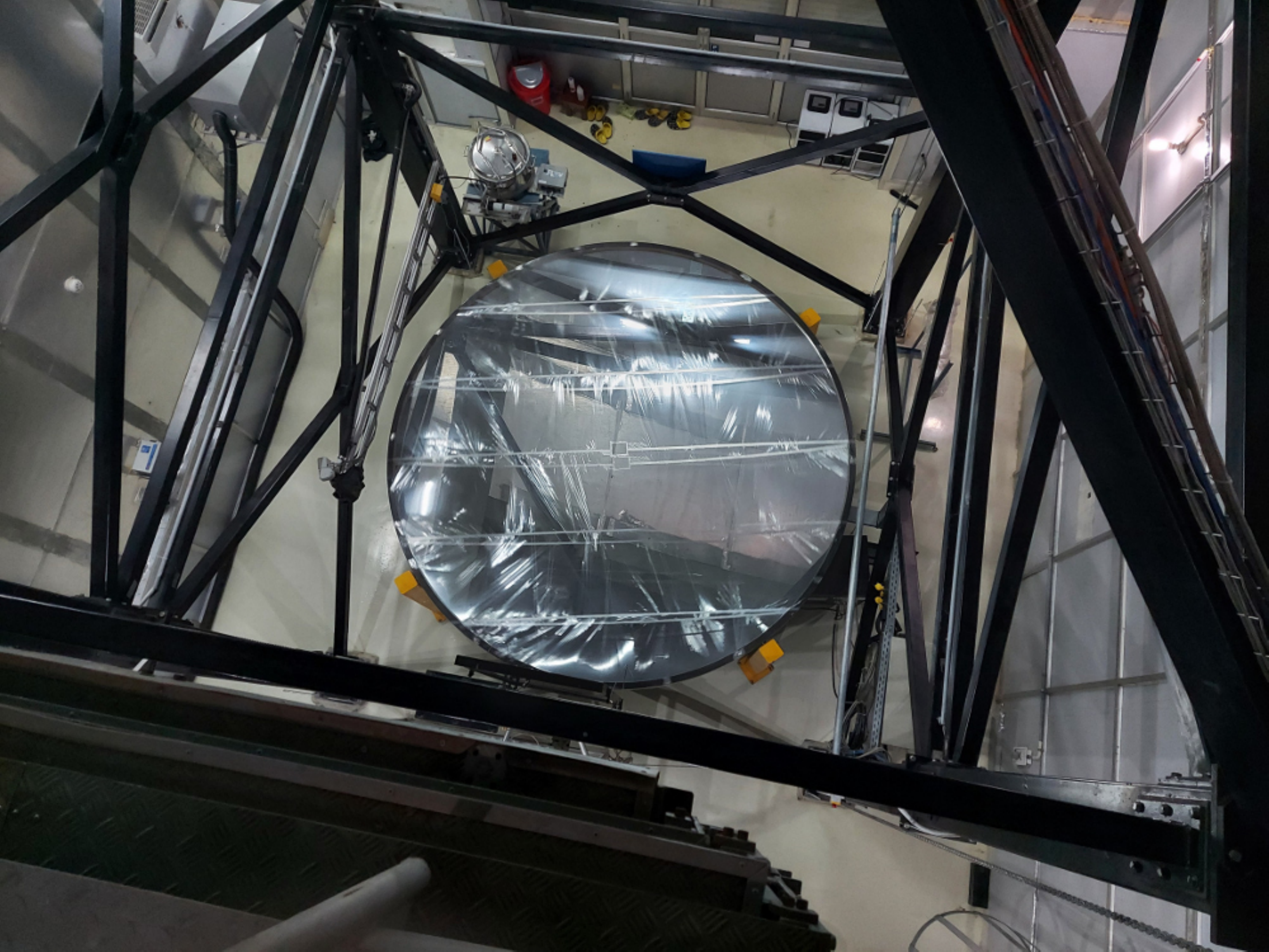}
\caption{Top view of the newly formed ILMT mirror. The adhesive tapes of the installed mylar cover are visible over the mirror.}
\label{ILMT-mirror}
\end{figure}

Finally, the mirror-formation activities were started. All the ventilators and hatch cover were opened for better ventilation in the ILMT main enclosure. Approximately 50 liters of mercury were transferred to the mirror from a storage tank by means of a mercury pumping system. This system includes a peristaltic pump, hoses, and a mechanism that lowers a tube into a depression at the centre of the mirror to allow mercury to be pumped on or off the mirror, even if it is rotating. The mirror-control system was then engaged to rotate the mirror at its nominal angular rate. 

Because of its high surface tension, the mercury does not flow uniformly and cover the surface. Instead, large areas of the mirror remain uncovered. These holes need to be filled by imparting small accelerations and decelerations to the mirror by means of a hand-held push-button control. After a duration of about 10 minutes, all holes were filled (see Fig.~\ref{ILMT-mirror}). The average thickness of the resulting mercury film was $\sim 3.8$ mm (determined by measuring the volume of mercury transferred to the mirror and dividing by the wetted area of the mirror).

During this entire process mercury safety procedures were strictly enforced, with only essential personnel in the room and all wearing appropriate personal protective equipment (PPE), including half-face respirators equipped with mercury-vapour cartridges. Nevertheless, continuous monitoring with two different kinds of mercury vapour detectors (model: VM-3000 and MVI) showed that vapour levels in the room and around the mirror never rose above 25 $\mu$g m$^{-3}$, the threshold level above which mercury-vapour respirators are needed. Both detectors are based on UV-absorption principle (wavelength 254 nm) and the measurement can be done at different ranges e.g. VM--3000: 1--100, 1--1000 and 1--2000 $\mu$g m$^{-3}$; MVI: 0--200 and 0-2000 $\mu$g m$^{-3}$ (at different resolution). Inside the ILMT enclosure, the mean value of vapour counts was 11 $\mu$g m$^{-3}$ ($\sigma$=5 $\mu$g m$^{-3}$). It subsequently reduced below mean value of 4 $\mu$g m$^{-3}$ on the next day.    

Along with the above, several other activities, such as checking the pneumatic modules, CCD camera cooling, filter slide movement, etc. were also completed to avoid any possible issue during the on-sky tests.

\begin{figure}
\centering
\includegraphics[width=\columnwidth]{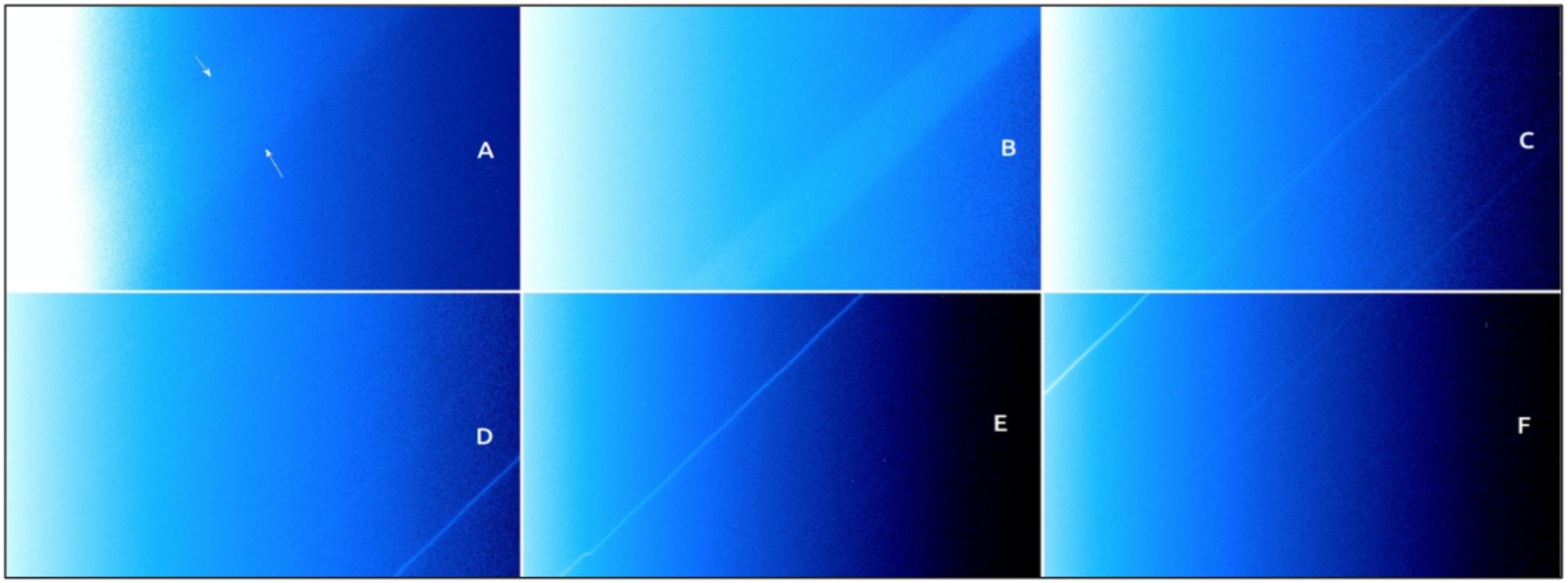}
\caption{Image trails for different focus values. For a better visibility, the image contrast is changed. The approximate trail width in panel A is indicated between the two arrows.}
\label{ILMT-FL-1}
\end{figure}

\section{First light observations}

Experience with previous liquid mirrors indicates that within two hours of startup, the focal length of the mirror stabilizes and the image quality improves greatly. This coincides with a drop in the rate of evaporation of the mercury, indicating that a thin film of oxide has formed on the mercury surface \citep{Hickson-2007LZT}. The ILMT first light observations were conducted on 2022 April 29 under moderate sky conditions (after six hours of mirror formation). Initially, multiple test frames were obtained and subsequently, we could see a very extended star trail (Fig.~\ref{ILMT-FL-1}, panel A). Then we tried to better focus the images while changing the vertical position of the corrector-CCD assembly which resulted in sharper image trails (Fig.~\ref{ILMT-FL-1}, panels B, C, D, E and F). A significant gradient of light was seen on the frames, most likely due to some light leakage in the dome. Therefore, all possible light emitting sources (e.g. LED indicators of the instruments) were blocked by pasting black tape or covering it with thick cloths. Finally, the light gradient from the images disappeared after switching off the mirror monitoring CCTV cameras.

It was realized that optimal focus was not within the range of the focus mechanism, so the focal length was slightly increased by adjusting the rotation speed of the mirror, i.e. setting a more appropriate frequency on the signal generator that provides the frequency reference for the control system. Next, the CCD columns were aligned along the E-W direction by adjusting the azimuth position of the CCD camera.

Multiple test images were obtained on successive nights, which allowed optimization of the azimuth of the CCD camera, the TDI scan rate and the focus. The latter depends on temperature, due to thermal expansion of the telescope structure. Thus, an empirical relation was determined between the optimal focus setting and the temperatures indicated by sensors attached to the telescope.

\begin{figure}
\centering
\includegraphics[width=\columnwidth]{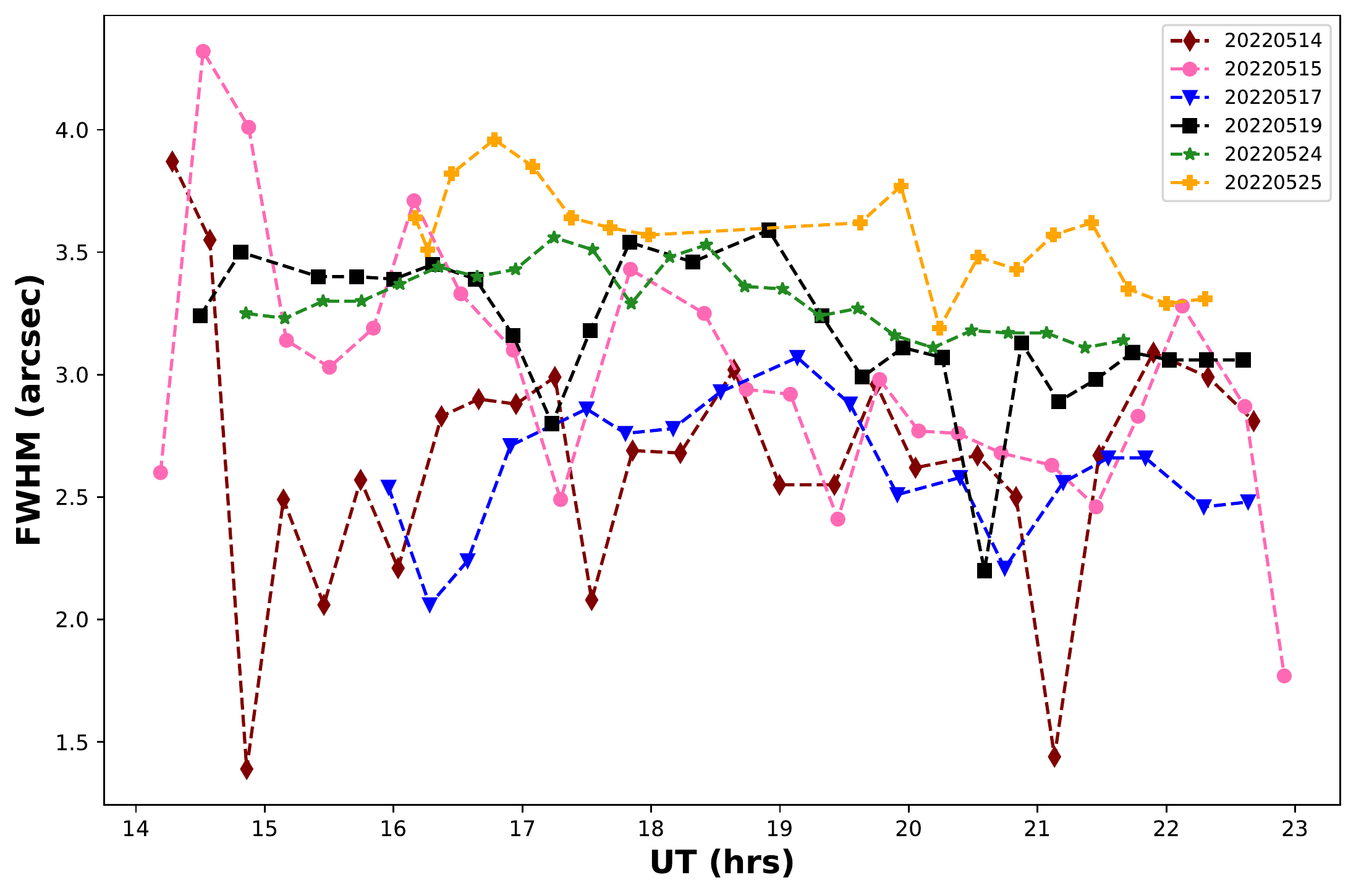}
\caption{Variation of the FWHM of the ILMT images taken on different nights is displayed.}
\label{ILMT-fwhm}
\end{figure}

\begin{figure}
\centering
\includegraphics[width=\columnwidth]{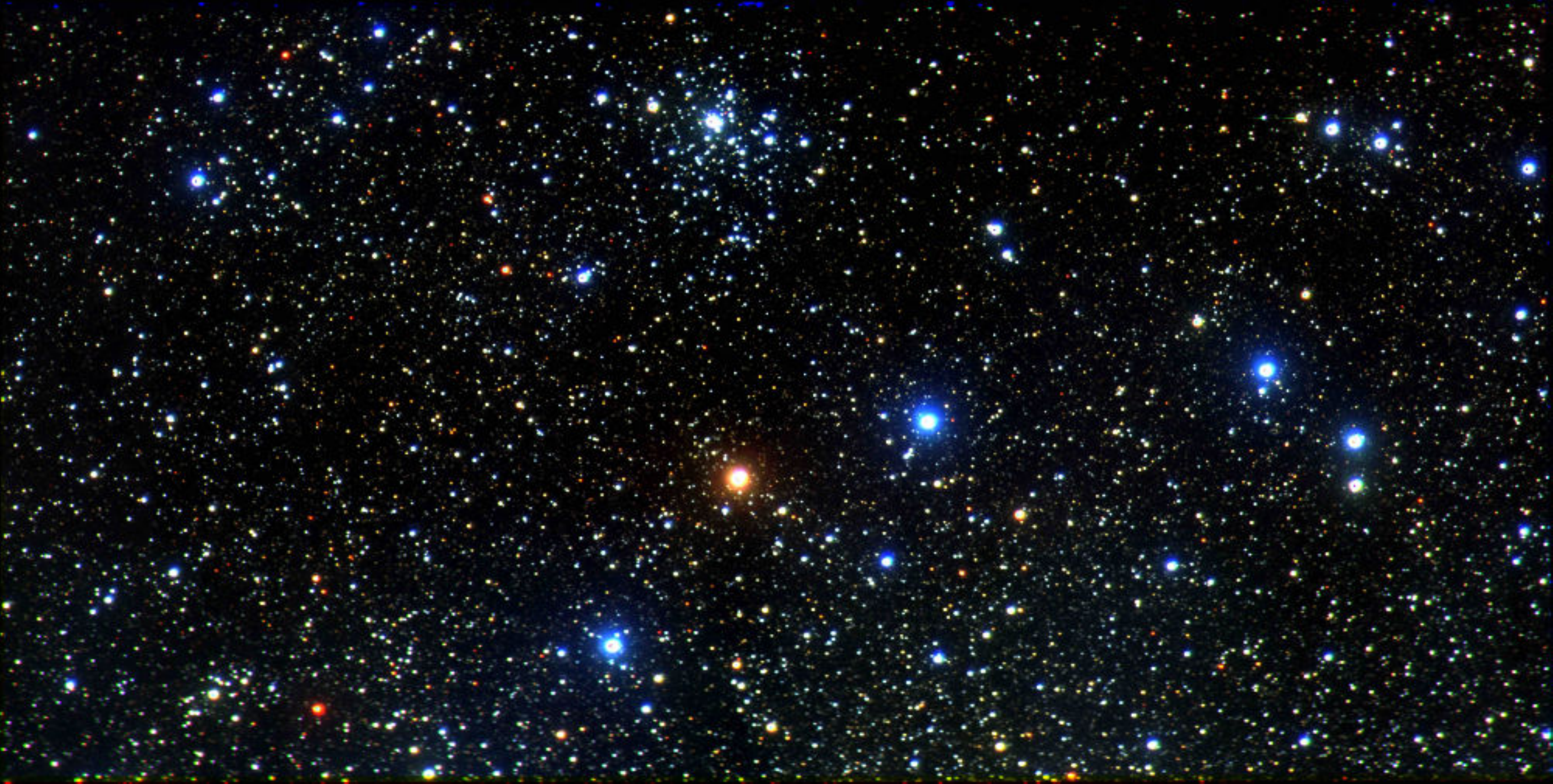}
\caption{Colour composite image of a strip of sky taken with the ILMT. This RGB image was made using single-scanned $g'$, $r'$ and $i'$ band images observed on different nights.}
\label{ILMT-image}
\end{figure}

An automated data reduction pipeline will be implemented to handle the data acquired with the ILMT. The pipeline includes pre-processing (e.g., dark subtraction, flat-field correction, and sky subtraction) and post-processing (astrometric and photometric calibration). However, at this stage, various modules and commands are executed manually. We examined the stellar profile of different night images after pre-processing them. The full width at half maximum (FWHM) was estimated for six nights. The variation of this parameter is shown in Fig.~\ref{ILMT-fwhm}, which indicates that generally, the FWHM ranges between 2-4 arcsec. Here, we emphasize that some abnormal elongations of the stellar images have been noticed towards the outer edges in the CCD frames (along declination). The astrometric and photometric investigation of the images is in progress (see Dukiya et al. in this issue). A colour-composite image, created from observations on three different nights, is shown in Fig.~\ref{ILMT-image}.

\section{Summary and work ahead}

The ILMT has recently seen first light at the ARIES Devasthal observatory in India. Crucial preparatory activities were executed before the first light, as discussed in Section~\ref{prep}. Commissioning-phase observations were conducted for several weeks before the facility had to be closed due to the Monsoon. 
Investigation of six night ILMT images indicates that the stellar FWHM varies between 2-4 arcsec. Further, analysis of the ILMT data is underway and also development and testing of the data reduction pipeline is planned. The upcoming ILMT survey will be particularly useful for scientific projects that deal with astrometric and/or photometric variability. Peculiar targets discovered with the ILMT can be followed-up with larger observing facilities.

It is worth noting that image quality of the ILMT depends critically on many factors such as leveling of the air bearing to within $\sim 1$ arcsec, leveling of the corrector to within 2 mrad, and its correct orientation with respect to the N-S direction (typically within one degree), accurate centering (x-y) of the optical corrector with respect to the mirror center, etc. Therefore, verification of these parameters is essential. The analysis of ILMT images indicates that stellar profiles near the central region were sharp and circular. However, the outer region images were affected with the TDI distortions. A raytrace analysis was performed which confirmed that the corrector had been installed with the wrong orientation. This issue will be resolved, during the monsoon shutdown, by rotating the optical corrector by 180$^{\circ}$ about the vertical axis. 
 
\section*{Acknowledgements}
The 4m International Liquid Mirror Telescope (ILMT) project results from a collaboration between Aryabhatta Research Institute of Observational Sciences (ARIES, India), the Institute of Astrophysics and Geophysics (University of Li\`ege, Belgium), the Canadian Astronomical Institutes, University of Montreal, University of Toronto, York University, University of British Columbia and Victoria University.
PH acknowledges financial support from the Natural Sciences and Engineering Research Council of Canada, RGPIN-2019-04369.

\bibliographystyle{ws-jai.bst}
\setlength{\bibsep}{0pt plus 0.3ex}
{\footnotesize
\bibliography{ILMT}
}
\nocite{*}
\end{document}